\documentclass[12pt,twoside,a4paper]{article}

\usepackage{qmc,psfig}

\newcommand{\be}{\begin{equation}}                                              
\newcommand{\ee}{\end{equation}}                                                
\newcommand{\half}{\frac{1}{2}}

\newcommand{\LCB}{\raisebox{-0.3ex}{\mbox{\LARGE$\left\{\right.$}}}
\newcommand{\RCB}{\raisebox{-0.3ex}{\mbox{\LARGE$\left.\right\}$}}}

\begin{document}

\title{ Majorana fermions on the lattice }

\author{ Istv\'an Montvay\footnote{Lecture given at the Workshop
``Quantum Monte Carlo'', ECT$^*$ Trento, July 2001.} } 

\instit{ Deutsches Elektronen-Synchrotron DESY,  \\
         Notkestr.\,85, D-22603 Hamburg, Germany }

\gdef\theauthor{ I.\ Montvay }
\gdef\thetitle{ Majorana fermions on the lattice }

\maketitle

\begin{abstract}
 The Monte Carlo simulation of Majorana fermions is discussed on the
 example of supersymmetric Yang-Mills (SYM) theory.
\end{abstract}

\section{Introduction}\label{sec1}
 Majorana fermions play an important r\^ole in supersymmetric quantum
 field theories which are the favorite candidates for the extension
 of the Standard Model of elementary particle interactions beyond the
 presently available energy range.
 It is generally assumed that the scale where supersymmetry becomes
 manifest is near to the presently explored electroweak scale and that
 the supersymmetry breaking is spontaneous.
 An attractive possibility for spontaneous supersymmetry breaking is to
 exploit non-perturbative mechanisms in supersymmetric gauge theories.

 Non-perturbative features of supersymmetric theories can be derived 
 analytically, as for instance the basic work of Seiberg and Witten
 \cite{SEIWIT} shows, and can also be obtained by numerical Monte Carlo
 simulations on a lattice.
 The numerical approach requires the introduction of Majorana
 fermion fields on a four-dimensional Euclidean lattice in space 
 and imaginary time.

 The simplest supersymmetric gauge theory is the supersymmetric
 extension of Yang-Mills gauge theory.
 It is the gauge theory of a massless Majorana fermion, called
 ``gaugino'', in the adjoint representation of the gauge group.
 The Euclidean action density of a gauge theory in the adjoint
 representation can be written as
\be  \label{eq01}
{\cal L}\; = \;\frac{1}{4}\, F^a_{\mu\nu}F^a_{\mu\nu}
+ \half\, \overline{\lambda}^a \gamma_\mu \left({\cal D}_\mu 
\lambda\right)^a + m_{\tilde{g}}\, \overline{\lambda}^a \lambda^a \ .
\ee
 Here $F^a_{\mu\nu}$ denotes the field strength tensor and $\lambda^a$
 is the Grassmannian fermion field, both with the adjoint representation
 index $a$.
 $m_{\tilde{g}}$ is the gaugino mass which has to be set equal to zero
 for supersymmetry.
 For a Majorana fermion $\lambda^a$ and $\overline{\lambda}^a$
 are not independent but satisfy
\be\label{eq02}
\overline{\lambda} = \lambda^T C \ ,
\ee
 with $C$ the charge conjugation Dirac matrix.
 This definition is based on the analytic continuation of Green's
 functions from Minkowski to Euclidean space \cite{MAJORANA}.

\section{Lattice formulation}\label{sec2}

 In order to define the path integral for a Yang Mills theory with
 Majorana fermions in the adjoint representation, let us first
 consider the familiar case of Dirac fermions \cite{MM}.
 (For a general reference on lattice quantum field theory see this
 book.)
 Let us denote the Grassmanian fermion fields in the adjoint
 representation by $\psi^a_x$ and $\overline{\psi}^a_x$.
 Here Dirac spinor indices are omitted for simplicity and $a$ stands for
 the adjoint representation index ($a=1,..,N_c^2-1$ for SU($N_c$) ).
 The fermionic part of the Wilson lattice action is 
\begin{eqnarray}  \label{eq03}
S_f &=& \sum_x \LCB \overline{\psi}_x^a\psi_x^a
\nonumber \\
 &-& K \sum_{\mu=1}^4 \left[
\overline{\psi}_{x+\hat{\mu}}^a V_{ab,x\mu}(1+\gamma_\mu)\psi_x^b
+\overline{\psi}_x^a V_{ab,x\mu}^T (1-\gamma_\mu)
\psi_{x+\hat{\mu}}^b \right] \RCB \ .
\end{eqnarray}
 Here $K$ is the hopping parameter, the Wilson parameter removing
 the fermion doublers in the continuum limit is fixed to $r=1$ and
 the matrix for the gauge-field link in the adjoint representation
 $V_{x\mu}$ is defined from the fundamental link variables $U_{x\mu}$
 according to
\be  \label{eq04}
V_{ab,x\mu} \equiv V_{ab,x\mu}[U] \equiv
2 {\rm Tr}(U_{x\mu}^\dagger T_a U_{x\mu} T_b)
= V_{ab,x\mu}^* =V_{ab,x\mu}^{-1T} \ .
\ee
 The generators $T_a \equiv \half \lambda_a$ satisfy the usual
 normalization ${\rm Tr\,}(\lambda_a\lambda_b)=\half$.
 In the simplest case of SU(2) ($N_c=2$) we have, of course,
 $T_a \equiv \half \tau_a$ with the isospin Pauli-matrices $\tau_a$.
 The normalization of the fermion fields in (\ref{eq03}) is the
 usual one for numerical simulations.
 The full lattice action is the sum of the pure gauge part and
 fermionic part: 
\be  \label{eq05}
S = S_g + S_f \ .
\ee
 The standard Wilson action for the SU($N_c$) gauge field $S_g$
 is a sum over the plaquettes
\be  \label{eq06}
S_g  =   \beta \sum_{pl}                                                  
\left( 1 - \frac{1}{N_c} {\rm Re\,Tr\,} U_{pl} \right) \ ,   
\ee
 with the bare gauge coupling given by $\beta \equiv 2N_c/g^2$.

 In order to obtain the lattice formulation of a theory with
 Majorana fermions let us note that out of a Dirac fermion field it
 is possible to construct two Majorana fields:
\be  \label{eq07}
\lambda^{(1)} \equiv \frac{1}{\sqrt{2}} ( \psi + C\overline{\psi}^T)
\ , \hspace{2em}
\lambda^{(2)} \equiv \frac{i}{\sqrt{2}} (-\psi + C\overline{\psi}^T)
\ee
 with the charge conjugation matrix $C$.
 These satisfy the Majorana condition
\be  \label{eq08}
\overline{\lambda}^{(j)} = \lambda^{(j)T} C
\hspace{3em} (j=1,2)  \ . 
\ee
 The inverse relation of (\ref{eq07}) is
\be  \label{eq09}
\psi = \frac{1}{\sqrt{2}} (\lambda^{(1)} + i\lambda^{(2)})
\ , \hspace{2em}
\psi_c \equiv C\overline{\psi}^T =
\frac{1}{\sqrt{2}} (\lambda^{(1)} - i\lambda^{(2)}) \ .
\ee
 In terms of the two Majorana fields the fermion action $S_f$ in
 eq.\ (\ref{eq03}) can be written as
\begin{eqnarray}  \label{eq10}
S_f &=& \half \sum_x \sum_{j=1}^2 \LCB
\overline{\lambda}_x^{(j)a}\lambda_x^{(j)a}
\nonumber \\
 &-& K \sum_{\mu=1}^4 \left[
\overline{\lambda}_{x+\hat{\mu}}^{(j)a} V_{ab,x\mu}
(1+\gamma_\mu)\lambda_x^{(j)b}
+\overline{\lambda}_x^{(j)a} V_{ab,x\mu}^T (1-\gamma_\mu)
\lambda_{x+\hat{\mu}}^{(j)b} \right] \RCB \ .
\end{eqnarray}

 For later purposes it is convenient to introduce the {\em fermion
 matrix}
\begin{eqnarray}  \label{eq11}
Q_{yd,xc} &\equiv& Q_{yd,xc}[U] \equiv \delta_{yx}\delta_{dc}
\nonumber \\
 &-& K \sum_{\mu=1}^4 \left[
\delta_{y,x+\hat{\mu}}(1+\gamma_\mu) V_{dc,x\mu} +
\delta_{y+\hat{\mu},x}(1-\gamma_\mu) V^T_{dc,y\mu} \right] \ .
\end{eqnarray}
 Here, as usual, $\hat{\mu}$ denotes the unit vector in direction
 $\mu$.
 In terms of $Q$ we have
\be  \label{eq12}
S_f = \sum_{xc,yd} \overline{\psi}^d_y Q_{yd,xc} \psi^c_x
= \half\sum_{j=1}^2
\sum_{xc,yd} \overline{\lambda}^{(j)d}_y Q_{yd,xc} \lambda^{(j)c}_x \ ,
\ee
 and the fermionic path integral can be written as
\be  \label{eq13}
\int [d\overline{\psi} d\psi] e^{-S_f} = 
\int [d\overline{\psi} d\psi] e^{-\overline{\psi} Q \psi} = \det Q
= \prod_{j=1}^2 \int [d\lambda^{(j)}] 
e^{ -\half\overline{\lambda}^{(j)}Q\lambda^{(j)} } \ .
\ee
 This shows that the path integral over the Dirac fermion is the square
 of the path integral over the Majorana fermion and therefore
\be  \label{eq14}
\int [d\lambda] e^{ -\half\overline{\lambda} Q \lambda }
= \pm \sqrt{\det Q} \ .
\ee
 As one can see here, for Majorana fields the path integral involves
 only $[d\lambda^{(j)}]$ because of the Majorana condition in
 (\ref{eq08}).

 The relation (\ref{eq14}) leaves the sign on the righ hand side
 undetermined.
 A unique definition of the path integral over a Majoran fermion field,
 including the sign, is given by
\be  \label{eq15}
\int [d\lambda] e^{ -\half\overline{\lambda} Q \lambda } 
= \int [d\lambda] e^{ -\half\lambda M \lambda } = {\rm Pf}(M)
\ee
 where $M$ is the antisymmetric matrix defined as
\be  \label{eq16}
M \equiv CQ = -M^T \ .
\ee

 The square root of the determinant in eq.\ (\ref{eq14}) is a
 {\em Pfaffian} \cite{PFAFFIAN}.
 This can be defined for a general complex antisymmetric matrix
 $M_{\alpha\beta}=-M_{\beta\alpha}$ with an even number of dimensions
 ($1 \leq \alpha,\beta \leq 2N$) by a Grassmann integral as
\be  \label{eq17}
{\rm Pf}(M) \equiv
\int [d\phi] e^{-\half\phi_\alpha M_{\alpha\beta} \phi_\beta}
= \frac{1}{N! 2^N} \epsilon_{\alpha_1\beta_1 \ldots \alpha_N\beta_N}
M_{\alpha_1\beta_1} \ldots M_{\alpha_N\beta_N} \ .
\ee
 Here, of course, $[d\phi] \equiv d\phi_{2N} \ldots d\phi_1$, and 
 $\epsilon$ is the totally antisymmetric unit tensor.

 It is now clear that the fermion action for a Majorana fermion in the
 adjoint representation $\lambda^a_x$ can be defined by
\begin{eqnarray}  \label{eq18}
S_f &\equiv& \half \overline{\lambda} Q \lambda \equiv 
\half \sum_x \LCB \overline{\lambda}_x^a\lambda_x^a
\nonumber \\
 &-& K \sum_{\mu=1}^4 \left[
\overline{\lambda}_{x+\hat{\mu}}^a V_{ab,x\mu}(1+\gamma_\mu)\lambda_x^b
+\overline{\lambda}_x^r V_{ab,x\mu}^T (1-\gamma_\mu)
\lambda_{x+\hat{\mu}}^b \right] \RCB \ .
\end{eqnarray}
 This together with (\ref{eq05})-(\ref{eq06}) gives a lattice action
 for the gauge theory of Majorana fermion in the adjoint representation.
 In order to achieve supersymmetry one has to tune the hopping parameter
 (bare mass parameter) $K$ to the {\em critical value} $K_{cr}(\beta)$
 in such a way that the mass of the fermion becomes zero.

 The path integral over $\lambda$ is defined by the Pfaffian
 ${\rm Pf}(CQ) = {\rm Pf}(M)$.
 By this definition the sign on the right hand side of 
 eq.\ (\ref{eq14}) is uniquely determined.
 The determinant $\det(Q)$ is real because the fermion matrix in
 (\ref{eq11}) satisfies
\be \label{eq19}
Q^\dagger = \gamma_5 Q \gamma_5 \ ,
\hspace{3em}
\tilde{Q} \equiv \gamma_5 Q = \tilde{Q}^\dagger \ .
\ee
 Moreover one can prove that $\det(Q)=\det(\tilde{Q})$ is always
 non-negative.
 This follows from the relations
\be\label{eq20}
CQC^{-1} = Q^T \ , \hspace{1em} B\tilde{Q}B^{-1} = \tilde{Q}^T \ ,
\ee
 with the charge conjugation matrix $C$ and $B \equiv C\gamma_5$.
 It follows that every eigenvalue of $Q$ and $\tilde{Q}$ is (at least)
 doubly degenerate.
 Therefore, with the real eigenvalues $\tilde{\lambda}_i$ of the
 Hermitean fermion matrix $\tilde{Q}$, we have
\be\label{eq21}
\det(Q) = \det(\tilde{Q}) = \prod_i \tilde{\lambda}_i^2 \geq 0 .
\ee
 Since according to the above discussion
\be\label{eq22}
\det(Q) = \det(M) = \left[ {\rm Pf}(M) \right]^2 \ ,
\ee
 the Pfaffian ${\rm Pf}(M)$ has to be real -- but it can have any sign.

\section{Numerical simulations of SYM theories}\label{sec3}

 In order to perform Monte Carlo simulations of SYM theory one needs
 a positive measure on the gauge field which allows for importance
 sampling of the path integral.
 Therefore the sign of the Pfaffian can only be taken into account by
 reweighting.
 According to (\ref{eq22}) the absolute value of the Pfaffian is the
 non-negative square root of the determinant therefore the effective
 gauge field action is \cite{CURVEN}:
\be \label{eq23}
S_{CV} = \beta\sum_{pl} \left( 1-\half{\rm Tr\,}U_{pl} \right)
- \half\log\det Q[U] \ .
\ee
 The factor $\half$ in front of $\log\det Q$ shows that we effectively
 have a flavour number $N_f=\half$ of adjoint fermions.
 The omitted sign of the Pfaffian can be taken into account by reweighting
 the expectation values according to
\be\label{eq24}
\langle A \rangle = \frac{\langle A\; {\rm sign Pf}(M)\rangle_{CV}}
{\langle {\rm sign Pf}(M)\rangle_{CV}} \ ,
\ee
 where $\langle \ldots \rangle_{CV}$ denotes expectation values with
 respect to the effective gauge action $S_{CV}$.
 This may give rise to a {\em sign problem} which will be discussed in
 section \ref{sec3.1}.

 The fractional power of the determinant corresponding to (\ref{eq23})
 can be reproduced, for instance, by the hybrid molecular dynamics
 algorithm \cite{HMD} which is, however, a finite step size algorithm
 where the step size has to be extrapolated to zero.
 An ``exact'' algorithm where the step size extrapolation is absent
 is the {\em two-step multi-bosonic} (TSMB) algorithm
 \cite{TSMB,POLYNOM}.
 The first large scale numerical simulation of SYM theory has recently
 been performed by the DESY-M\"unster-Roma collaboration using the
 TSMB algorithm \cite{DISCHIRAL,SPECTRUM,WTI}.

\subsection{The ``sign problem''}\label{sec3.1}

 The Pfaffian resulting from the Grassmannian path integrals for
 Majorana fermions (\ref{eq15}) is an object similar to a determinant
 but less often used.
 As shown by (\ref{eq17}), ${\rm Pf}(M)$ is a polynomial of the
 matrix elements of the $2N$-dimensional antisymmetric matrix $M=-M^T$.
 Basic relations are \cite{PFAFFIAN}
\be\label{eq25}
M = P^T J P ,\hspace{3em}  {\rm Pf}(M)=\det(P) \ ,
\ee
 where $J$ is a block-diagonal matrix containing on the diagonal
 $2\otimes2$ blocks equal to $\epsilon=i\sigma_2$ and otherwise zeros.
 Let us note that from these relations the second equality in
 eq.~(\ref{eq22}) immediately follows.

 The form of $M$ required in (\ref{eq25}) can be achieved by a
 procedure analogous to the Gram-Schmidt orthogonalization and, by
 construction, $P$ is a triangular matrix (see \cite{SPECTRUM}).
 This gives a numerical procedure for the computation of $P$ and the
 determinant of $P$ gives, according to (\ref{eq25}), the Pfaffian
 ${\rm Pf}(M)$.
 Since $P$ is triangular, the calculation of $\det(P)$ is, of course,
 trivial.

 This procedure can be used for a numerical determination of the
 Pfaffian on small lattices.
 On lattices larger than, say, $4^3 \cdot 8$ the computation becomes
 cumbersome due to the large storage requirements.
 This is because one has to store a full $\Omega \otimes \Omega$ matrix,
 with $\Omega$ being the number of lattice points multiplied by the
 number of spinor-colour indices (equal to $4(N_c^2-1)$ for the adjoint
 representation of ${\rm SU}(N_c)$).
 The difficulty of computation is similar to a computation of the
 determinant of $Q$ with $LU$-decomposition.

 Fortunately, in order to obtain the sign of the Pfaffian occurring in
 the reweighting formula (\ref{eq24}), one can proceed without a full
 calculation of the value of the Pfaffian.
 The method is to monitor the sign changes of ${\rm Pf}(M)$ as a
 function of the hopping parameter $K$.
 According to (\ref{eq21}), the hermitean fermion matrix for the gaugino
 $\tilde{Q}$ has doubly degenerate real eigenvalues therefore
\be\label{eq26}
\det M = \det \tilde{Q} = \prod_{i=1}^{\Omega/2} \tilde{\lambda}_i^2 \ ,
\ee
 where $\tilde{\lambda}_i$ denotes the eigenvalues of $\tilde{Q}$.
 This implies
\be\label{eq27}
|{\rm Pf}(M)| =  \prod_{i=1}^{\Omega/2} |\tilde{\lambda}_i| \ ,
\hspace{2em} \Longrightarrow \hspace{2em}
{\rm Pf}(M) =  \prod_{i=1}^{\Omega/2} \tilde{\lambda}_i \ .
\ee
 The first equality trivially follows from (\ref{eq22}).
 The second one is the consequence of the fact that ${\rm Pf}(M)$ is
 a polynomial in $K$ which cannot have discontinuities in any of its
 derivatives.
 Therefore if, as a function of $K$, an eigenvalue $\tilde{\lambda}_i$
 (or any odd number of them) changes sign the sign of ${\rm Pf}(M)$
 has to change, too.
 Since at $K=0$ we have ${\rm Pf}(M)=1$, the number of sign changes
 between $K=0$ and the actual value of $K$, where the dynamical fermion
 simulation is performed, determines the sign of ${\rm Pf}(M)$.
 This means that one has to determined the flow of the eigenvalues of
 $\tilde{Q}$ through zero \cite{SPECTRALFLOW}.

 The spectral flow method is well suited for the calculation of the
 sign of the Pfaffian in SYM theory.
 An important question is the frequency and the effects of
 configurations with negative sign.
 A strongly fluctuating Pfaffian sign is a potential danger for the
 effectiveness of the Monte Carlo simulation because cancellations
 can occur resulting in an unacceptable increase of statistical errors.
 The experience of the DESY-M\"unster Collaboration shows, however,
 that below the critical line $K_{cr}(\beta)$ corresponding to zero
 gaugino mass ($m_{\tilde{g}}=0$) negative Pfaffians practically never
 appear \cite{SPECTRUM,WTI}.
 Above the critical line several configurations with negative Pfaffian
 have been observed but their r\^ole has not yet been cleared up to
 now.
 Since supersymmetry is expected to be realized in the continuum limit
 at  $m_{\tilde{g}}=0$, the negative signs of the Pfaffian can be
 avoided if one takes the zero gaugino mass limit from
 $m_{\tilde{g}} > 0$ corresponding to $K < K_{cr}$.
 In this sense there is no ``sign problem'' in SYM which would
 prevente a Monte Carlo investigation.

 The presence or absence of negative Pfaffians in a sample of gauge
 configurations produced in Monte Carlo simulations can be easily
 seen even without the application of the spectral flow method.
 In case of sign changes the distribution of the smallest eigenvalues
 of the squared fermion matrix $\tilde{Q}^2$ shows a pronounced tail
 reaching down to zero \cite{DESYSWANSEA}.
 The absence of a tail shows that there are no negative Pfaffians.

 Concerning this ``sign problem'' let us note that a very similar
 phenomenon appears also in QCD because the Wilson-Dirac determinant of
 a single quark flavour can also have a negative sign.
 Under certain circumstances the sign of the quark determinant plays
 an important r\^ole.
 This is the case, for instance, at large quark chemical potential in a
 QCD-like model with SU(2) colour and staggered quarks in the adjoint
 representation which has recently been studied by the DESY-Swansea
 Collaboration \cite{DESYSWANSEA}.
 This investigation also revealed an interesting feature of the TSMB
 algorithm, namely its ability to easily change the sign of eigenvalues
 of the hermitean fermion matrix (and hence the sign of the determinant
 or Pfaffian).
 This is in contrast to algorithms based on finite difference molecular
 dynamics equations as, for instance, the HMD \cite{HMD} algorithms.

\section{Outlook}\label{sec4}

 Numerical simulations of $N=1$ supersymmetric theories require to deal
 with Majorana fermions on a Euclidean lattice.
 Such simulations are feasible with presently available computer
 technology and using well established simulation algorithms -- at
 least in the relatively simple case of supersymmetric Yang-Mills
 theories \cite{DISCHIRAL,SPECTRUM,WTI,DGHV,FKV}.
 Many interesting questions are waiting for detailed answers.
 Just to mention a few: the behaviour of the phase transition at the
 supersymmetric point, the spectroscopy of supersymmetric multiplets in
 the particle spectrum and the Ward-Takahashi identities proving the
 realization of supersymmetry in quantum field theories.

 {\bf Acknowledgement:} It is a great pleasure to thank the members of
 the DESY-M\"unster-Roma collaboration for our pleasant common work
 on supersymmetric Yang-Mills theories.


\end{document}